\documentstyle[psfig,twocolumn,prl,aps,amssymb]{revtex}
\begin{document}
\wideabs{
\draft

\date{\today} \title{The excitonic collapse of higher Landau level fractional quantum
Hall effect}
\author{V.W. Scarola, Kwon Park, and J.K. Jain}
\address{Department of Physics, 104 Davey Laboratory, The Pennsylvania State
University, University Park, Pennsylvania 16802}
\maketitle

\begin{abstract}

The scarcity of the fractional quantum
Hall effect in higher Landau levels is a most intriguing fact when
contrasted with its great abundance in the lowest Landau level. This paper shows that
a suppression of the hard core repulsion in going from the lowest Landau level to
higher Landau levels leads to a collapse of the energy of the
neutral excitation, destabilizing all fractional states in
the third and higher Landau levels, and almost all in the second Landau level.  
The remaining fractions are in agreement with those observed experimentally.

\end{abstract}

}

Electrons in the
lowest Landau level exhibit the spectacular phenomenon of the fractional quantum
Hall effect (FQHE).\cite{Tsui}  Including the several new fractions observed in the ultra-low
temperature measurements in very high quality samples,\cite{Pan}
there now exists evidence for more than 50 fractions in the lowest Landau level.
The essential phenomenon is remarkably insensitive to the detailed
form of the repulsive interaction, as manifested by the fact that it is
quite robust to perturbations arising from
Landau level mixing, finite thickness, or the nature of the transverse confinement.

The system of electrons restricted to a higher Landau
level (LL) deviates from that in the lowest LL
only through the short-distance matrix elements for the Coulomb interaction.
Consequently, one would expect the FQHE to be not too sensitive to the Landau
level index either.  From this point of view, it is astounding that the fractional quantum
Hall effect (FQHE) is so rare in higher Landau levels.  For example, whereas 10 members of
each of the sequences $\nu=n/(2n+1)$ and $\nu=n/(2n-1)$
have been observed in the lowest LL, only the first one
has been observed in the second LL, and none at all in the third and higher LLs.  
In fact, the only 
fractions outside the lowest LL for which clear experimental evidence exists, in the form of reasonably
well quantized plateaus, are $\nu=1/3$ and $\nu=1/2$ in the second LL.\cite{Pan,fh}
(That the latter has no analog in the {\em lowest} LL further underscores the
striking difference between the lowest and the higher LL physics.)
Hartree-Fock variational (Koulakov, Fogler, and Shklovskii\cite {Fogler}),
exact diagonalization (Rezayi, Haldane, and Yang \cite{Rezayi}), and experimental 
(Lilly {\em et al.}, Du {\em et al.} \cite {anisotropy})
studies make a compelling case that a bubble crystal or an 
anisotropic stripe phase is favored over the FQHE in higher LLs.

The goal of this work is to start from the FQHE end of the problem and 
ask by what mechanism is the FQHE destroyed in
higher Landau levels.  Of course, there can be a mundane origin for the
disappearance of the FQHE, e.g., sufficiently strong disorder, but of interest
to us here is the possible {\em intrinsic} instability of FQHE in higher
Landau levels.  To this end, we will start by assuming an incompressible
FQHE state and then investigate its stability to quantum fluctuations.
A necessary condition for FQHE is that (in the absence of disorder) the
energy gap to all excitations remain positive
definite.  A vanishing of the gap signals an
instability of the assumed ground state.  It will be
shown that the suppression of the short-distance Coulomb matrix elements in
higher Landau levels leads to a
collapse of the energy of the neutral exciton, destabilizing
most of the incompressible FQHE states in higher LLs.

The investigations will be carried out within the framework of the composite
fermion (CF) theory\cite{Jain} of the FQHE, which
has proven successful in capturing subtle instabilities of the FQHE.
At small filling factors
($\nu\leq 1/9$), the neutral composite-fermion exciton becomes gapless.\cite{Kamilla}
In another example, a recent theoretical study \cite{Cooper}
has revealed that the Fermi sea of composite fermions is unstable to Cooper pairing
at $\nu=5/2$ but not at $\nu=1/2$, giving insight into the experimental observation of
the FQHE at $\nu=5/2$ but not at $\nu=1/2$.\cite {fh,Pan}

A composite fermion, $^{2p}$CF, is the bound state of an electron and $2p$
quantum mechanical vortices.  Of relevance to
experiment are three flavors of composite fermions carrying two, four, and six
vortices, namely $^2$CFs, $^4$CFs, and $^6$CFs.  According to
the composite fermion theory, the interacting electrons at the LL filling factor
$\nu_{0}\equiv \frac{n}{2pn+1}$ transform into weakly interacting composite
fermions with an effective filling $\nu^{*}_{0}=n$.  The ground state
here corresponds to $n$ filled CF-LLs and the neutral excitation to a particle-hole
pair of composite fermions, called the CF exciton.
The explicit, parameter-free, lowest-LL form for the microscopic wave functions
for the fully polarized CF ground state and the CF exciton
can be found in the literature\cite{JK} and will not be repeated here.
In the lowest Landau level, these wave functions have been found to be
accurate in tests against
exact diagonalization results available for small systems.\cite{Jain,Review1}
They are not as good in higher Landau levels for $^2$CFs, but should still
be a reasonable starting point {\em provided} the ground state is an incompressible
state.  Interestingly, the wave functions are quantitatively accurate for $^4$CFs in the
second LL.\cite {AR}

We will study filling factors $\nu=2s+\nu_{0}$.  Here, the quantity $s=0,1,...$
denotes the Landau level index, with
the lowest $s$ LLs having {\em both} spin states filled, thus contributing
$2s$ to the filling factor.  As a simplification
we will take these $2s$ LLs to be completely inert (neglecting by the lower Landau 
levels\cite{Aleiner}) and work only with the
electrons in the topmost partially filled LL, which will be taken to be fully spin
polarized; this is a valid approximation in the
limit of high magnetic fields when LL mixing is negligible.  Note that only the
electrons in the topmost partially filled LL capture vortices to
turn into composite fermions; the electrons in the lower, filled LLs remain unchanged.

\begin{figure}
\centerline{\psfig{figure=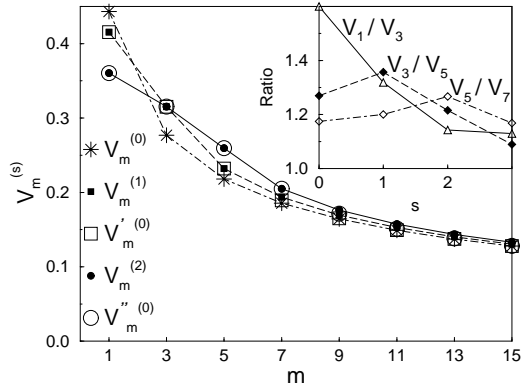,height=2.5in,angle=-90}}
\caption{The Coulomb pseudopotentials for the lowest three Landau levels,
$s=0$ (stars) $s=1$ (filled squares), and $s=2$ (filled circles).
Also shown are the pseudopotentials of the effective interactions $V'(r)$ and $V''(r)$
in the lowest LL (empty squares and empty circles).  The inset shows the ratios
$V_1/V_3$, $V_3/V_5$, and $V_5/V_7$ in various Landau
levels. The lines are a guide to the eye.}
\label{fig1}
\end{figure}

The wave functions $\Psi$ are most easily constructed within the lowest LL.
To calculate energies in higher LLs one may promote them to
higher LLs by application of the Landau level raising operator.  However, this
procedure is technically rather cumbersome.  We instead proceed by working
with an {\em effective} interaction in the lowest LL which mimics the
Coulomb interaction in a higher LL.
The interparticle interaction in any given Landau level is completely specified by
its Haldane pseudopotentials, $V^{(s)}_{m}$, defined  via
$V(r_i-r_j)=\sum_m V^{(s)}_{m} P^{(s)ij}_{m},$
where $V^{(s)}_{m}$ is the interaction energy of two particles in the $s$th LL in
the relative angular momentum $m$ state, and $P^{(s)ij}_{m}$ is the corresponding
projection operator.\cite{Haldane}
Following Park {\em et al.},\cite{Park2} we
map the problem of a given interaction in the $s=1$  Landau level
into that of an {\em effective} interaction $V'(r)$ in the lowest
($s=0$) Landau level, chosen so that the two have the same pseudopotentials, i.e.,
$V^{(1)}_{m}=V'^{(0)}_{m}$.
We implement this strategy in an approximate scheme
by taking the following convenient form for the effective potential:
$$
V'(r) = \frac{e^2}{\epsilon}
\left(\frac{1}{r} + a'_1 e^{-\alpha'_1 r^2} + a'_2 r^2e^{-\alpha'_2 r^2}
\right)\;.
$$
The parameters $a'_1$, $a'_2$, $\alpha'_1$, and $\alpha'_2$ are fixed by
requiring that the first three to four {\em odd} pseudopotentials of the effective potential
in the lowest LL match exactly the corresponding pseudopotentials of the
Coulomb potential in the second Landau level.
Only the odd pseudopotentials are relevant due to the antisymmetric form of the spatial
part of the wave function.  The remaining higher order pseudopotentials are
asymptotically correct because $V'(r)\rightarrow \frac{e^2}{\epsilon r}$
for large $r$.  The $s=2$ LL can be similarly treated, and the corresponding
effective potential will be denoted by $V''(r)$.  The pseudopotentials for the Coulomb
interaction in the $s=1$ and $s=2$ LLs and for the corresponding
effective potentials $V'(r)$ and $V''(r)$ in the $s=0$ LL are shown
in Fig.~(\ref{fig1}), demonstrating that the lowest LL problem with the
effective interaction is an excellent approximation to the higher LL problem with Coulomb
interaction.
With this interaction, we then evaluate the energy of the CF exciton using the quantum
Monte Carlo method developed earlier for the lowest LL wave functions.\cite{JK}
Because the pseudopotential are matched for the planar geometry, our results in the
spherical geometry are meaningful only for sufficiently large systems; unphysical
results may be obtained for small systems due to finite size effects.
We will work below with systems containing as many as $N=66$ particles, using
an efficient updating method discussed earlier.\cite{Scarola}

\begin{figure}
\centerline{\psfig{figure=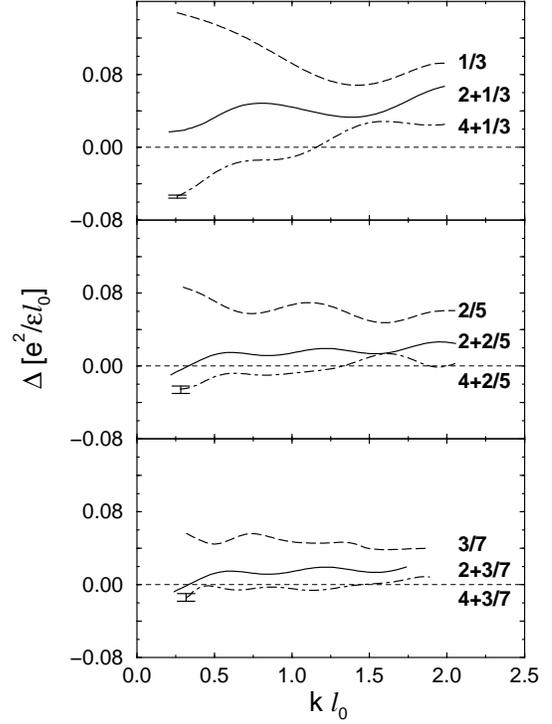,height=4in,angle=0}}
\caption{The dispersions of the composite-fermion exciton
in the lowest three Landau levels: $s=0$ (dashed line), $s=1$ (solid line), and $s=2$ (dot-
dashed line) at $\nu=2s+\nu_0$, with $\nu_0=1/3$, 2/5, and 3/7.  The typical Monte Carlo
uncertainty is shown at the beginning of the $s=1$ curves.  The energies are given in
units of $e^{2}/\epsilon\l_0$ where $\epsilon$ is the dielectric constant of the background
material, and $l_{0}$ is the magnetic length at $\nu$.
}
\label{fig2}
\end{figure}

We first consider $^2$CFs, corresponding to FQHE at $\nu=2s+\frac{n}{2n + 1}$.
Fig.~(\ref{fig2}) shows the energy of the CF exciton with $n=1$, 2, and 3 in
the three lowest LLs ($s=0$, 1, and 2), for 66 particles.  The energies
are quoted in units of $e^2/\epsilon\l_0$, where $l_0=\sqrt{\hbar c/eB}$ is
the magnetic length at $\nu$.

All FQHE is unstable in $s=2$  for $^2$CFs.  We expect that this would remain
the case in still higher LLs.
The absence of FQHE in $s\geq 2$ is consistent with earlier theoretical
studies,\cite {Fogler,Rezayi,Sbeouelji}
which have made a convincing case for either a
bubble crystal or a stripe phase in third or higher ($s\geq 2$) LLs.
At half filling ($\nu_0=1/2$), the wave vector of the stripe phase was 
estimated\cite{Fogler,Rezayi}
to be $q l_0\sim 2.4/\sqrt{2s+1}$, which is also approximately equal to 
the reciprocal lattice vector associated with the bubble crystal away from half filling. 
Even though the instability occurs for wave vectors below $ql_0\approx 1$, it 
is unfortunately not possible to determine from the dispersions shown in
Fig.~(\ref{fig2}) a {\em single} wave vector for the instability, which precludes us
from ascertaining from our method the
reciprocal lattice vector of the true charge density wave ground state.
Note that a comparison between the energies of the 1/3 FQHE state
and the Wigner crystal state in the third LL does not
indicate a lack of FQHE here,\cite {GM} which is understandable in view of the
fact that the instability is into a more complicated bubble crystal.

Surprisingly, we find the FQHE to be unstable also in the second LL. 
The only exception is $\nu_0=1/3$, which corresponds to $\nu=7/3$ in experiments 
(and, of course, other states related to it by symmetry).  
We have confirmed that $\nu=7/3$ remains stable in the thermodynamic limit by
extrapolating the $kl_0\rightarrow 0$ energy to the $N^{-1}\rightarrow 0$ limit.
Given that, by its very design, our approach is expected to
{\em overestimate} the strength of a FQHE state, the results provide strong evidence against FQHE 
at 12/5 for the pure Coulomb interaction.  There is often a minimum observed in $\rho_{xx}$ in the vicinity of
$\nu=12/5$,\cite {Pan,Eisenstein} which may suggest an incipient FQHE
state here (although no plateau has been observed yet).
Can a FQHE state here be stabilized by LL mixing or finite thickness effects?
LL mixing is a weak effect at large magnetic fields, and 
is expected to have the opposite effect, because
it effectively screens the short range part of the interaction, thereby
further weakening a given FQHE state.  (It is possible, however, that at relatively
low magnetic fields, when the screening is strong, the modification of the
interaction due to screening by the lower Landau level\cite{Aleiner} might alter this conclusion.
We have not investigated this question.)  We have considered the effect of
the transverse width of the electron wave function by employing the Fang-Howard
variational form for the effective interaction,\cite{Fang} appropriate for
the triangular well geometry.  The instability is weakened, but not eliminated 
for typical paramters.

The instability in the second Landau level 
(Fig.~\ref{fig2}) appears to occur at a small wave vector,
which suggests a uniform compressible ground state (although
a charge density wave state with a large lattice spacing can
obviously not be ruled out on account
of the finite size of our study).  The nature of the compressible state in the 
second LL is not fully understood at present.

\begin{figure}
\centerline{
\psfig{figure=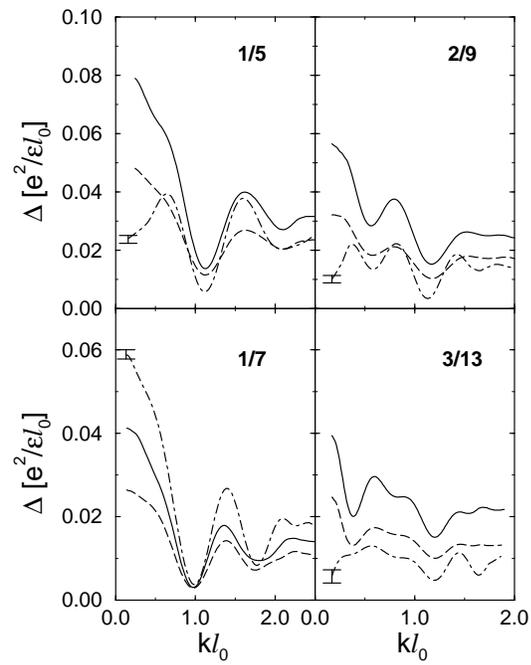,height=4in,angle=0}}
\caption{The same as Fig.~(2) but for the fractions with
$\nu_0=1/5$, 2/9, 3/13, and 1/7.}
\label{fig3}
\end{figure}

What about the other flavors of composite fermions?
There are theoretical indications \cite{GM} that these are
more stable in higher LLs than the $^2$CFs.
In order to explore this issue further, we have computed the dispersion
of the CF exciton for the $^4$CFs and
$^6$CFs at $\nu=2s+1/5, 2s+2/9, 2s+3/13$ and $2s+1/7$
for $s=0,$ 1, and 2, shown
in Fig.~(\ref{fig3}).  Indeed, these states are
more stable in higher Landau levels, with the $^4$CF states having the largest
roton gap (in units of $e^2/\epsilon l_0$) in the second LL and
the $^6$CF states in the third.
Unlike for $^2$CFs, FQHE for $^4$CFs and $^6$CFs
survives in the second and third LLs.  As mentioned earlier, the trial 
wave function $\Psi$ is quite close to the actual 1/5 ground state in the second LL,
in fact more accurate than in the lowest LL,\cite{AR}
implying that the dispersion shown in Fig.~(\ref{fig3}) is quantitatively reliable. The
2/7 state is analogous to 1/5:  whereas 1/5 is obtained from 1/3 by attachment of
two additional vortices, 2/7 is similarly obtained from 2/3.
There exists preliminary experimental evidence for both 1/5 and 2/7 in the second
LL.\cite{Pan,Eisenstein}  There is an indication for 1/7 state in the lowest
LL \cite {Goldman} but none yet in higher LLs.  The observation of the $^4$CF
and $^6$CF states in higher LLs is complicated by their
rather small energy gaps (because, for a given density,  $e^2/\epsilon l_0$ is
much smaller here than at
the corresponding lowest LL fraction), as well as their close proximity to
strong integral quantum Hall plateaus.

Why is the FQHE for $^2$CFs unstable in higher Landau levels but relatively
stable to changes in effective interaction within the lowest LL?  The reason is that
in order for the FQHE to occur, the interaction must not only be repulsive but
it must have a sufficiently strong hard-core repulsion at short distances.
It is useful to characterize the hard-core nature of the repulsion
for a given interaction by the ratio $V^{(s)}_1/V^{(s)}_3$, given in Fig.~(\ref{fig1}).
It appears that for $V_1/V_3\leq 1.3$, most FQHE is
unstable, with the exception of the 1/3 state which
is only marginally stable.  In contrast, when finite
thickness is taken into account in the lowest LL, all pseudopotentials are
suppressed more or less uniformly, and the
ratio remains above 1.4-1.5 for typical experimental parameters.
For $^4$CFs the pseudopotential $V_1$ is ineffective  (provided it is not too small to cause
an instability) because the wave functions for $^4$CFs are approximately given by
the $^2$CF wave functions multiplied by a power of Jastrow factor that
eliminates the unit relative angular momentum.
Similarly, $V_1$ and $V_3$ are not relevant for $^6$CFs.  Therefore,
the appropriate ratios are $V_3/V_5$ and $V_5/V_7$ for $^4$CFs and $^6$CFs,
respectively.  These, however,
{\em increase} in going from the lowest to the higher LLs, as shown in Fig.~(\ref{fig1}),
thus explaining why $^4$CFs and $^6$CFs have a qualitatively different LL-index dependence
as compared to the $^2$CFs.  There is a close correspondence between the appropriate
ratio and the stability, as a comparison of Figs.~(\ref{fig1}) and (\ref{fig3}) shows, with
$^4$CFs ($^6$CFs) being strongest in the $s=1$ ($s=2$) LL.

In conclusion, our study provides an insight into the paucity of
FQHE in higher Landau levels, as well as the qualitative difference between
the stabilities of composite fermions of various flavors in different Landau levels.
Most FQHE states of $^2$CFs in higher Landau levels are intrinsically
unstable due to a collapse of the neutral excitation, which in turn
is triggered by a softening of the short range part of the interaction.

This work was supported in part by the National Science Foundation under grant no.
DMR-9986806.  We are grateful to the Numerically Intensive Computing Group led by V.
Agarwala, J. Holmes, and J. Nucciarone, at the Penn State University CAC for
assistance and computing time with the LION-X cluster.
We thank J.P. Eisenstein, V.J. Goldman, E. Rezayi, and H.L. Stormer for helpful conversations.

\end{document}